\documentclass[reprint]{revtex4}%
\usepackage{amsfonts}
\usepackage{amsmath}
\usepackage{amssymb}
\usepackage{charter}
\usepackage{graphicx}%
\providecommand{\U}[1]{\protect \rule{.1in}{.1in}}
\begin{document}
\title{Statistical properties and decoherence of two-mode photon-subtracted squeezed
vacuum\thanks{{\small Work supported by the the National Natural Science
Foundation of China under Grant Nos.10775097 and 10874174.}}}
\author{Li-yun Hu$^{1,2}$\thanks{{\small Corresponding author. Email:
hlyun@sjtu.edu.cn; hlyun2008@126.com.}}, Xue-xiang Xu$^{1,2}$ and Hong-yi
Fan$^{2}$}
\affiliation{$^{1}${\small College of Physics and Communication Electronics, Jiangxi Normal
University, Nanchang 330022, China}}
\affiliation{$^{2}${\small Department of Physics, Shanghai Jiao Tong University, Shanghai,
200240, China}}

\begin{abstract}
{\small We investigate the statistical properties of the photon-subtractions
from the two-mode squeezed vacuum state and its decoherence in a thermal
environment. It is found that the state can be considered as a squeezed
two-variable Hermite polynomial excitation vacuum and the normalization of
this state is the Jacobi polynomial of the squeezing parameter. The compact
expression for Wigner function (WF) is also derived analytically by using the
Weyl ordered operators' invariance under similar transformations. Especially,
the nonclassicality is discussed in terms of the negativity of WF. The effect
of decoherence on this state is then discussed by deriving the analytical time
evolution results of WF. It is shown that the WF is always positive for any
squeezing parameter and photon-subtraction number if the decay time exceeds an
upper bound (}$\kappa t>\frac{1}{2}\ln \frac{2\bar{n}+2}{2\bar{n}+1}).$

\end{abstract}
\maketitle

Key Words: photon-subtraction; nonclassicality; Wigner function; negativity;
two-mode squeezed vacuum state

PACS numbers: 03.65.Yz, 42.50.Dv

\section{Introduction}

Entanglement is an important resource for quantum information
\cite{1}. In a quantum optics laboratory, Gaussian states, being
characteristic of Gaussian Wigner functions, have been generated but
there is some\ limitation in using them for various tasks of quantum
information procession \cite{2}. For example, in the first
demonstration of continuous variables quantum teleportation
(two-mode squeezed vacuum state as a quantum channel), the squeezing
is low, thus the entanglement of the quantum channel is such low
that the average fidelity of quantum teleportation is just more
$\left( 8\pm2\right)  \%$ than the classical limits. In order to
increase the quantum entanglement there have been suggestions and
realizations to engineering the quantum state by subtracting or
adding photons from/to a Gaussian field which are plausible ways to
conditionally manipulate nonclassical state of optical field
\cite{2a,2b,2c,2d,2e,2f,2g}. In fact, such methods allowed the
preparation and analysis of several states with negative Wigner
functions, including one- and two-photon Fock states \cite{3,4,5,6},
delocalized single photons \cite{7,8}, and photon-subtracted
squeezed vacuum states (PSSV), very similar to quantum
superpositions of coherent states with small amplitudes (a
Schr\"{o}dinger kitten state \cite{9,10,11,12}) for single-mode
case.

Recently, the two-mode PSSVs (TPSSVs) have been paid enough attention by both
experimentalists and theoreticians \cite{3,4,13,14,15,16,17,18,19,20,21,22}.
Olivares et al \cite{13,14} considered the photon subtraction using on--off
photodetectors and showed the improvement of quantum teleportation depending
on various parameters involved. Then they further studied the nonlocality of
photon subtraction state in the presence of noise \cite{15}. Kitagawa et al
\cite{16}, on the other hand, investigated the degree of entanglement for the
TPSSV by using an on-off photondetector. Using operation with single photon
counts, Ourjoumtsev et al .\cite{3,4} have demonstrated experimentally that
the entanglement between Gaussian entangled states, can be increased by
subtracting only one photon from two-mode squeezed vacuum states. The resulted
state is a complex quantum state with a negative two-mode Wigner function.
However, so far as we know, there is no report about\ the nonclassicality and
decoherence of TPSSV for arbitrary number PSSV in literature before.

In this paper, we will explore theoretically the statistical properties and
decoherence of arbitrary number TPSSV. This paper is arranged as follows: in
Sect. II we introduce the TPSSV, denoted as $a^{m}b^{n}S_{2}(\lambda
)\left \vert 00\right \rangle ,$ where $S_{2}(\lambda)$ is two-mode squeezing
operator with $\lambda$ being squeezing parameter and $m,n$ are the subtracted
photon number from $S_{2}(\lambda)\left \vert 00\right \rangle $ for mode $a$
and $b$, respectively. It is found that it is just a squeezed two-variable
Hermite polynomial excitation on the vacuum state, and then the normalization
factor for $a^{m}b^{n}S_{2}(\lambda)\left \vert 00\right \rangle $ is derived,
which turns out to be a Jacobi polynomial, a remarkable result. In Sec. III,
the quantum statistical properties of the TPSSV, such as distribution of
photon number, squeezing properties, cross-correlation function and
antibunching, are calculated analytically and then be discussed in details.
Especially, in Sec. IV, the explicit analytical expression of Wigner function
(WF) of the TPSSV is derived by using the Weyl ordered operators' invariance
under similar transformations, which is related to the two-variable
Gaussian-Hermite polynomials, and then its nonclassicality is discussed in
terms of the negativity of WF which implies the highly nonclassical properties
of quantum states. Sec. V is devoted to studying the effect of the decoherence
on the TPSSV in a thermal environment. The analytical expressions for the
time-evolution of the state and its WF are derived, and the loss of
nonclassicality is discussed in reference of the negativity of WF due to
decoherence. We find that the WF for TPSSV has no chance to present negative
value for all parameters $\lambda$ and $m,n$ if the decay time $\kappa
t>\frac{1}{2}\ln \frac{2\bar{n}+2}{2\bar{n}+1}\ $(see Eq.(\ref{47}) below),
where $\bar{n}$ denotes the average thermal photon number in the
environment\textbf{\ }with dissipative coefficient\textbf{\ }$\kappa$.

\section{Two-mode photon-subtracted squeezed vacuum states}

\subsection{TPSSV as the squeezed two-variable Hermite polynomial excitation
state}

The definition of the two-mode squeezed vacuum state is given by%
\begin{equation}
S_{2}(\lambda)\left \vert 00\right \rangle =\text{sech}\lambda \exp(a^{\dagger
}b^{\dagger}\tanh \lambda)\left \vert 00\right \rangle ,\label{1}%
\end{equation}
where $S_{2}(\lambda)=\exp[\lambda(a^{\dagger}b^{\dagger}-ab)]$ is the
two-mode squeezing operator \cite{23,24,25} with $\lambda$ being a real
squeezing parameter, and $a$, $b$ are the Bose annihilation operators,
$[a,a^{\dagger}]=[b,b^{\dagger}]=1$. Theoretically, the TPSSV can be obtained
by repeatedly operating the photon annihilation operator $a$ and $b$ on
$S_{2}(\lambda)\left \vert 00\right \rangle $,\ defined as%
\begin{equation}
\left \vert \lambda,m,n\right \rangle =a^{m}b^{n}S_{2}(\lambda)\left \vert
00\right \rangle ,\label{2}%
\end{equation}
where $\left \vert \lambda,m,n\right \rangle $ is an un-normalization state.
Noticing the transform relations,%
\begin{align}
S_{2}^{\dagger}(\lambda)aS_{2}(\lambda) &  =a\cosh \lambda+b^{\dagger}%
\sinh \lambda,\nonumber \\
S_{2}^{\dagger}(\lambda)bS_{2}(\lambda) &  =b\cosh \lambda+a^{\dagger}%
\sinh \lambda,\label{3}%
\end{align}
we can reform Eq.(\ref{2}) as
\begin{align}
\left \vert \lambda,m,n\right \rangle  &  =S_{2}(\lambda)S_{2}^{\dagger}%
(\lambda)a^{m}b^{n}S_{2}(\lambda)\left \vert 00\right \rangle \nonumber \\
&  =S_{2}(\lambda)(a\cosh \lambda+b^{\dagger}\sinh \lambda)^{m}(b\cosh
\lambda+a^{\dagger}\sinh \lambda)^{n}\left \vert 00\right \rangle \nonumber \\
&  =S_{2}(\lambda)\sinh^{n+m}\lambda \sum_{l=0}^{m}\frac{m!\coth^{l}\lambda
}{l!\left(  m-l\right)  !}b^{\dagger m-l}a^{l}a^{\dagger n}\left \vert
00\right \rangle .\label{4}%
\end{align}
Further note that $a^{\dagger n}\left \vert 0\right \rangle =\sqrt{n!}\left \vert
n\right \rangle $ and $a^{l}\left \vert n\right \rangle =\frac{\sqrt{n!}}%
{\sqrt{(n-l)!}}\left \vert n-l\right \rangle =\frac{\sqrt{n!}}{(n-l)!}a^{\dagger
n-l}\left \vert 0\right \rangle $, leading to $a^{l}a^{\dagger n}\left \vert
00\right \rangle =\frac{n!}{(n-l)!}a^{\dagger n-l}\left \vert 00\right \rangle $,
thus Eq.(\ref{4}) can be re-expressed as%
\begin{align}
\left \vert \lambda,m,n\right \rangle  &  =S_{2}(\lambda)\sinh^{n+m}\lambda
\sum_{l=0}^{\min \left(  m,n\right)  }\frac{m!n!\coth^{l}\lambda}{l!\left(
m-l\right)  !(n-l)!}a^{\dagger n-l}b^{\dagger m-l}\left \vert 00\right \rangle
\nonumber \\
&  =\frac{\sinh^{\left(  n+m\right)  /2}2\lambda}{\left(  i\sqrt{2}\right)
^{n+m}}S_{2}(\lambda)\sum_{l=0}^{\min(m,n)}\frac{(-1)^{l}n!m!\left(
i\sqrt{\tanh \lambda}b^{\dagger}\right)  ^{m-l}\left(  i\sqrt{\tanh \lambda
}a^{\dagger}\right)  ^{n-l}}{l!(n-l)!\left(  m-l\right)  !}\left \vert
00\right \rangle \nonumber \\
&  =\frac{\sinh^{\left(  n+m\right)  /2}2\lambda}{\left(  i\sqrt{2}\right)
^{n+m}}S_{2}(\lambda)H_{m,n}\left(  i\sqrt{\tanh \lambda}b^{\dagger}%
,i\sqrt{\tanh \lambda}a^{\dagger}\right)  \left \vert 00\right \rangle ,\label{5}%
\end{align}
where in the last step we have used the definition of the two variables
Hermitian polynomials \cite{26,27}, i.e.,
\begin{equation}
H_{m,n}\left(  \epsilon,\varepsilon \right)  =\sum_{k=0}^{\min \left(
m,n\right)  }\frac{\left(  -1\right)  ^{k}m!n!\epsilon^{m-k}\varepsilon^{n-k}%
}{k!(m-k)!(n-k)!}.\label{6}%
\end{equation}
From Eq.(\ref{5}) one can see clearly that the TPSSV $\left \vert
\lambda,m,n\right \rangle $ is equivalent to a two-mode squeezed two-variable
Hermite-excited vacuum state and exhibits the exchanging symmetry, namely,
interchanging $m\Leftrightarrow n$ is equivalent to $a^{\dagger}%
\Leftrightarrow b^{\dagger}$. It is clear that, when $m=n=0,$ Eq.(\ref{5})
just reduces to the two-mode squeezed vacuum state due to $H_{0,0}=1$; while
for $n\neq0$ and $m=0,$ noticing $H_{0,n}\left(  x,y\right)  =y^{n},$
Eq.(\ref{5}) becomes ($N_{\lambda,0,n}^{-1}=n!\sinh^{2n}\lambda)$, see
Eq.(\ref{11}) below) $\left \vert \lambda,0,n\right \rangle =S_{2}%
(\lambda)\left \vert n,0\right \rangle ,$which is just a squeezed number state,
corresponding to a pure negative binomial state \cite{28}.

\subsection{The normalization of $\left \vert \lambda,m,n\right \rangle $}

To know the normalization factor $N_{\lambda,m,n}$ of $\left \vert
\lambda,m,n\right \rangle $, let us first calculate the overlap $\left \langle
\lambda,m+s,n+t\right.  \left \vert \lambda,m,n\right \rangle $. For this
purpose, using the first equation in Eq.(\ref{5}) one can express $\left \vert
\lambda,m,n\right \rangle $ as%
\begin{equation}
\left \vert \lambda,m,n\right \rangle =S_{2}(\lambda)\sum_{l=0}^{\min(m,n)}%
\frac{m!n!\sinh^{n+m}\lambda \coth^{l}\lambda}{l!\sqrt{(m-l)!(n-l)!}}\left \vert
n-l,m-l\right \rangle ,\label{7}%
\end{equation}
which leading to
\begin{align}
&  \left \langle \lambda,m+s,n+t\right.  \left \vert \lambda,m,n\right \rangle
\nonumber \\
&  =m!\left(  n+s\right)  !\delta_{s,t}\sinh^{2n+2m+2s}\lambda \nonumber \\
&  \times \sum_{l=0}^{\min(m,n)}\frac{\left(  m+s\right)  !n!\coth
^{2l+s}\lambda}{l!(m-l)!(n-l)!(l+s)!},\label{8}%
\end{align}
where $\delta_{s,t}$ is the Kronecker delta function. Without lossing the
generality, supposing $m<n$ and comparing Eq.(\ref{8}) with the standard
expression of Jacobi polynomials \cite{29}
\begin{equation}
P_{m}^{(\alpha,\beta)}(x)=\left(  \frac{x-1}{2}\right)  ^{m}\sum_{k=0}%
^{m}\left(
\begin{array}
[c]{c}%
m+\alpha \\
k
\end{array}
\right)  \left(
\begin{array}
[c]{c}%
m+\beta \\
m-k
\end{array}
\right)  \left(  \frac{x+1}{x-1}\right)  ^{k},\label{9}%
\end{equation}
one can put Eq.(\ref{8}) into the following form
\begin{equation}
\left \langle \lambda,m+s,n+t\right.  \left \vert \lambda,m,n\right \rangle
=m!\left(  n+s\right)  !\delta_{s,t}\sinh^{2n+s}\lambda \cosh^{s}\lambda
P_{m}^{(n-m,s)}(\allowbreak \cosh2\lambda),\label{10}%
\end{equation}
which is just related to Jacobi polynomials. In particular, when $s=t=0$, the
normalization constant $\emph{N}_{m,n,\lambda}$ for the state $\left \vert
\lambda,m,n\right \rangle $ is given by%
\begin{equation}
N_{\lambda,m,n}=\left \langle \lambda,m,n\right.  \left \vert \lambda
,m,n\right \rangle =m!n!\sinh^{2n}\lambda P_{m}^{(n-m,0)}(\cosh2\lambda
),\label{11}%
\end{equation}
which is important for further studying analytically the statistical
properties of the TPSSV. For the case $m=n$, it becomes Legendre polynomial of
the squeezing parameter $\lambda$, because of $P_{n}^{(0,0)}(x)=P_{n}(x),$
$P_{0}(x)=1$; while for $n\neq0$ and $m=0,$ noticing that $P_{0}^{(n,0)}(x)=1$
then $N_{\lambda,0,n}=n!\sinh^{2n}\lambda.$ Therefore, the normalized TPSSV
is
\begin{equation}
\left \vert \left \vert \lambda,m,n\right \rangle \right.  \equiv \left[
m!n!\sinh^{2n}\lambda P_{m}^{(n-m,0)}(\cosh2\lambda)\right]  ^{-1/2}a^{m}%
b^{n}S_{2}(\lambda)\left \vert 00\right \rangle .\label{12}%
\end{equation}

From Eqs. (\ref{2}) and (\ref{11}) we can easily calculate the average photon
number in TPSSV (denoting $\tau=\cosh2\lambda$),%

\begin{align}
\left \langle a^{\dagger}a\right \rangle  &  =N_{\lambda,m,n}^{2}\left \langle
00\right \vert S_{2}^{\dag}(\lambda)a^{\dag m+1}b^{\dag n}a^{m+1}b^{n}%
S_{2}(\lambda)\left \vert 00\right \rangle \nonumber \\
&  =\left(  m+1\right)  \frac{P_{m+1}^{(n-m-1,0)}(\tau)}{P_{m}^{(n-m,0)}%
(\tau)},\label{21}\\
\left \langle b^{\dagger}b\right \rangle  &  =\left(  n+1\right)  \sinh
^{2}\lambda \frac{P_{m}^{(n-m+1,0)}(\tau)}{P_{m}^{(n-m,0)}(\tau)}.\label{22}%
\end{align}
In a similar way we have%
\begin{equation}
\left \langle a^{\dagger}b^{\dagger}ab\right \rangle =\left(  m+1\right)
\left(  n+1\right)  \sinh^{2}\lambda \frac{P_{m+1}^{(n-m,0)}(\tau)}%
{P_{m}^{(n-m,0)}(\tau)}.\label{23}%
\end{equation}
Thus the cross-correlation function $g_{12}^{(2)}$ can be obtained by
\cite{30,31,32}
\begin{align}
g_{12}^{(2)}(\lambda) &  =\frac{\left \langle a^{\dagger}b^{\dagger
}ab\right \rangle }{\left \langle a^{\dagger}a\right \rangle \left \langle
b^{\dagger}b\right \rangle }\nonumber \\
&  =\frac{P_{m+1}^{(n-m,0)}(\tau)}{P_{m+1}^{(n-m-1,0)}(\tau)}\frac
{P_{m}^{(n-m,0)}(\tau)}{P_{m}^{(n-m+1,0)}(\tau)}.\label{20}%
\end{align}
Actually, the cross-correlation between the two modes reflects correlation
between photons in two different modes, which plays a key role in rendering
many two-mode radiations nonclassically. In Fig.1, we plot the graph of
$g_{12}^{(2)}\left(  \lambda \right)  $ as the function of $\lambda$ for some
different ($m,n$) values. It is shown that $g_{12}^{(2)}\left(  \lambda
\right)  $ are always larger than unit, thus there exist correlations between
the two modes. We emphasize that the WF has negative region for all $\lambda,$
and thus the TPSSV is nonclassical. In our following work, we pay attention to
the ideal TPSSV. \begin{figure}[ptb]
\label{Fig2}
\centering \includegraphics[width=14cm]{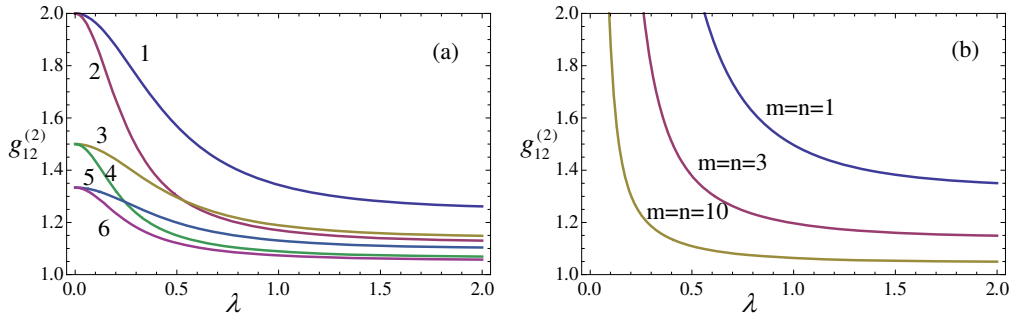}\caption{(Color online)
Cross-correlation function between the two modes $a$ and $b$ as a function of
$\lambda$ for different parameters ($m,n$). The number 1,2,3,4,5,6 in (a)
denote that ($m,n$) are eauql to (1,2), (3,4), (2,4),(6,8),(3,6) and (7,10)
respectively.}%
\end{figure}

\section{Quantum statistical properties of the TPSSV}

\subsection{\textit{Squeezing properties}}

For a two-mode system, the optical quadrature phase amplitudes can be
expressed as follows:
\begin{equation}
Q=\frac{Q_{1}+Q_{2}}{\sqrt{2}},\text{ }P=\frac{P_{1}+P_{2}}{\sqrt{2}},\text{
}[Q,P]=\mathtt{i}, \label{13}%
\end{equation}
where $Q_{1}=(a+a^{\dagger})/\sqrt{2}$, $P_{1}=(a-a^{\dagger})/(\sqrt
{2}\mathtt{i})$, $Q_{2}=(b+b^{\dagger})/\sqrt{2}$ and $P_{2}=(b-b^{\dagger
})/(\sqrt{2}\mathtt{i})$ are coordinate- and momentum- operator, respectively.
Their variances are $(\Delta Q)^{2}=\left \langle Q^{2}\right \rangle
-\left \langle Q\right \rangle ^{2}$ and $(\Delta P)^{2}=\left \langle
P^{2}\right \rangle -\left \langle P\right \rangle ^{2}$. The phase
amplifications satisfy the uncertainty relation of quantum mechanics $\Delta
Q\Delta P\geq1/2$. By using Eqs.(\ref{10}) and (\ref{11}), it is easy to see
that $\left \langle a\right \rangle =\left \langle a^{\dagger}\right \rangle
=\left \langle b\right \rangle =\left \langle b^{\dagger}\right \rangle =0\ $and
$\left \langle a^{2}\right \rangle =\left \langle a^{\dagger2}\right \rangle
=\left \langle b^{2}\right \rangle =\left \langle b^{\dagger2}\right \rangle =0$
as well as $\left \langle ab^{\dagger}\right \rangle =\left \langle a^{\dagger
}b\right \rangle =0,$ which leads to $\left \langle Q\right \rangle =0$,
$\left \langle P\right \rangle =0$. Moreover, using Eq.(\ref{10}) one can see
\begin{equation}
\left \langle a^{\dagger}b^{\dagger}\right \rangle =\left \langle ab\right \rangle
=\frac{n+1}{2}\frac{P_{m}^{(n-m,1)}(\tau)}{P_{m}^{(n-m,0)}(\tau)}\sinh
2\lambda. \label{14}%
\end{equation}
From Eqs.(\ref{21}), (\ref{22}) and (\ref{14}) it then follows that
\begin{align}
(\Delta Q)^{2}  &  =\frac{1}{2}(\left \langle a^{\dagger}a\right \rangle
+\left \langle b^{\dagger}b\right \rangle +\left \langle ab\right \rangle
+\left \langle a^{\dagger}b^{\dagger}\right \rangle +1)\nonumber \\
&  =\frac{1}{2P_{m}^{(n-m,0)}(\tau)}[\left(  m+1\right)  P_{m+1}%
^{(n-m-1,0)}(\tau)+\left(  n+1\right)  P_{m}^{(n-m+1,0)}(\tau)\sinh^{2}%
\lambda \nonumber \\
&  +\left(  n+1\right)  P_{m}^{(n-m,1)}(\tau)\sinh2\lambda+P_{m}%
^{(n-m,0)}(\tau)], \label{31}%
\end{align}
and
\begin{align}
(\Delta P)^{2}  &  =\frac{1}{2}(\left \langle a^{\dagger}a\right \rangle
+\left \langle b^{\dagger}b\right \rangle -\left \langle ab\right \rangle
-\left \langle a^{\dagger}b^{\dagger}\right \rangle +1)\nonumber \\
&  =\frac{1}{2P_{m}^{(n-m,0)}(\tau)}[\left(  m+1\right)  P_{m+1}%
^{(n-m-1,0)}(\tau)+\left(  n+1\right)  P_{m}^{(n-m+1,0)}(\tau)\sinh^{2}%
\lambda \nonumber \\
&  -\left(  n+1\right)  P_{m}^{(n-m,1)}(\tau)\sinh2\lambda+P_{m}%
^{(n-m,0)}(\tau)]. \label{32}%
\end{align}
Next, let us analyze some special cases. When $m=n=0,$ corresponding to the
two-mode squeezed state, Eqs. (\ref{31}) and (\ref{32}) becomes, respectively,
to
\begin{equation}
\left.  (\Delta Q)^{2}\right \vert _{m=n=0}=\frac{1}{2}\allowbreak e^{2\lambda
},\left.  (\Delta P)^{2}\right \vert _{m=n=0}=\frac{1}{2}\allowbreak
e^{-2\lambda},\Delta Q\Delta P=\frac{1}{2},
\end{equation}
which is just the standard squeezing case; while for $m=0,$ $n=1,$ Eqs.
(\ref{31}) and (\ref{32}) reduce to
\begin{equation}
\left.  (\Delta Q)^{2}\right \vert _{m=0,n=1}=\allowbreak e^{2\lambda},\left.
(\Delta P)^{2}\right \vert _{m=0,n=1}=\allowbreak \allowbreak e^{-2\lambda
},\Delta Q\Delta P=1, \label{33}%
\end{equation}
from which one can see that the state $\left \vert \left \vert \lambda
,m,n\right \rangle \right.  $ is squeezed at the \textquotedblleft p-direction"
when $\allowbreak \allowbreak e^{-2\lambda}<\frac{1}{2},$ i.e., $\lambda
>\frac{1}{2}\ln2$. In addition, when $m=n=1,$ in a similar way, one can get%
\begin{align}
\left.  (\Delta Q)^{2}\right \vert _{m=n=1}  &  =\frac{1}{2}e^{2\lambda}\left(
1+\frac{2e^{2\lambda}-2}{e^{2\lambda}+e^{-2\lambda}}\right)  ,\nonumber \\
\left.  (\Delta P)^{2}\right \vert _{m=n=1}  &  =\frac{1}{2}\frac
{1-e^{2\lambda}-3e^{-2\lambda}\left(  1-\allowbreak e^{-2\lambda}\right)
}{e^{2\lambda}+e^{-2\lambda}}+\frac{1}{2}<\frac{1}{2}, \label{7.43}%
\end{align}
which indicates that, for any squeezing parameter $\lambda$, there always
exist squeezing effect for state $|\left \vert \lambda,1,1\right \rangle $ at
the \textquotedblleft p-direction".

In order to see clearly the fluctuations of $(\Delta P)^{2}$ with other
parameters $m,n$ values, $\allowbreak$the figures are ploted in Fig.2. From
Fig.2(a) one can see that the fluctuations of $(\Delta P)^{2}$ are always less
than $\frac{1}{2}$ when $m=n,$ say, the state $|\left \vert \lambda
,m,m\right \rangle $ is always squeezed at the \textquotedblleft p-direction";
for given $m$ values, there exist the squeezing effect only when the squeezing
parameter exceeds a certain threshold value that increases with the
increasement of $n$ (see Fig.2(b)).

\begin{figure}[ptb]
\label{Fig8.1} \centering
\includegraphics[width=12cm]{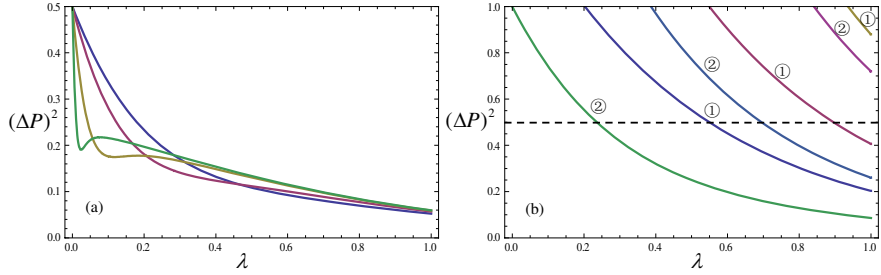}\caption{(Color online)
The fluctuations variation of $(\Delta P)^{2}$ with $\lambda$ for
several different parameters $m,n$ values: (a) $m=n=1,2,8,35$ from
down to up; (b) (1)
and (2) denote $m=0$ and $m=1$, respectively, and $n=2,5,12$ from down to up.}%
\end{figure}

\subsection{\textit{Distribution of photon number}}

In order to obtain the photon number distribution of the TPSSV, we begin with
evaluating the overlap between two-mode number state $\left \langle n_{a}%
,n_{b}\right \vert $ and $\left \vert \lambda,m,n\right \rangle .$ Using
Eq.(\ref{1}) and the un-normalized coherent state \cite{30,31}, $\left \vert
z\right \rangle =\exp \left(  za^{\dagger}\right)  \left \vert 0\right \rangle $,
leading to $\left \langle n\right \vert =\frac{1}{\sqrt{n!}}\left.
\frac{\partial^{n}}{\partial z^{\ast n}}\left \langle z\right \vert \right \vert
_{z^{\ast}=0}$, it is easy to see that%
\begin{align}
&  \left \langle n_{a},n_{b}\right.  \left \vert \lambda,m,n\right \rangle
\nonumber \\
&  =\text{sech}\lambda \left \langle n_{a},n_{b}\right \vert a^{m}b^{n}%
e^{a^{\dagger}b^{\dagger}\tanh \lambda}\left \vert 00\right \rangle \nonumber \\
&  =\frac{\left(  m+n_{a}\right)  !}{\sqrt{n_{a}!n_{b}!}}\text{sech}%
\lambda \tanh^{m+n_{a}}\lambda \delta_{m+n_{a},n+n_{b}}.\label{15}%
\end{align}
It is easy to follow that the photon number distribution of $|\left \vert
\lambda,m,n\right \rangle $, i.e.,
\begin{align}
P(n_{a},n_{b}) &  =N_{\lambda,m,n}^{-1}\left \vert \left \langle n_{a}%
,n_{b}\right.  \left \vert \lambda,m,n\right \rangle \right \vert ^{2}\nonumber \\
&  =\frac{\left[  \left(  m+n_{a}\right)  !\text{sech}\lambda \tanh^{m+n_{a}%
}\lambda \delta_{m+n_{a},n+n_{b}}\right]  ^{2}}{n_{a}!n_{b}!m!n!\sinh
^{2n}\lambda P_{m}^{(n-m,0)}(\cosh2\lambda)}.\label{18}%
\end{align}
From Eq.(\ref{18}) one can see that there exists a constrained condition,
$m+n_{a}=n+n_{b},$ for the photon number distribution (see Fig. 3). In
particular, when $m=n=0,$ Eq.(\ref{18}) becomes
\begin{equation}
P(n_{a},n_{b})=\left \{
\begin{array}
[c]{cc}%
\text{sech}^{2}\lambda \tanh^{2n_{a}}\lambda, & n_{a}=n_{b}\\
0, & n_{a}\neq n_{b}%
\end{array}
\right.  ,\label{19}%
\end{equation}
which is just the photon number distribution (PND) of two-mode squeezed vacuum state.

In Fig. 3, we plot the distribution $P(n_{a},n_{b})$ in the Fock space
($n_{a},n_{b}$) for some given $m,n$ values and squeezing parameter $\lambda$.
From Fig. 3 it is found that the PND is constrained by $m+n_{a}=n+n_{b},$
resulting from the paired-present of photons in two-mode squeezed state. By
subtracting photons, we have been able to move the peak from zero photons to
nonzero photons (see Fig.3 (a) and (c)). The position of peak depends on how
many photons are annihilated and how much the state is squeezed initially. In
addition, for example, the PND mainly shifts to the bigger number states and
becomes more \textquotedblleft flat" and \textquotedblleft wide" with the
increasing parameter $\lambda$ (see Fig.3 (b) and (c)).

\begin{figure}[ptb]
\label{Fig8.2}
\centering \includegraphics[width=12cm]{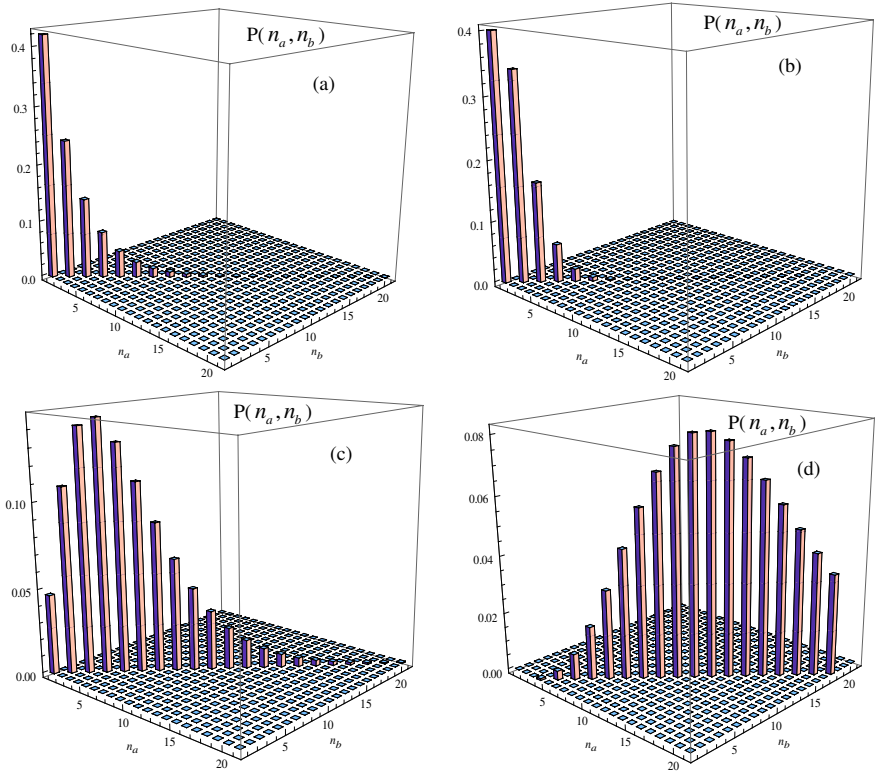}\caption{(Color online)
Photon number distribution $P(n_{a},n_{b})$ in the Fock space ($n_{a},n_{b}$)
for some given $m=n$ values: (a) $m=n=0,$ $\lambda=1$, (b) $m=n=1,$
$\lambda=0.5,$(c) $m=n=1,$ $\lambda=1,$(d) $m=2,n=5,$ $\lambda=1.$}%
\end{figure}

\subsection{\textit{Antibunching effect of the TPSSV}}

Next we will discuss the antibunching for the TPSSV. The criterion for the
existence of antibunching in two-mode radiation is given by \cite{33}%
\begin{equation}
R_{ab}\equiv \frac{\left \langle a^{\dagger2}a^{2}\right \rangle +\left \langle
b^{\dagger2}b^{2}\right \rangle }{2\left \langle a^{\dagger}ab^{\dagger
}b\right \rangle }-1<0.\label{25}%
\end{equation}
In a similar way to Eq.(\ref{21}) we have%
\begin{equation}
\left \langle a^{\dagger2}a^{2}\right \rangle =\left(  m+1\right)  \left(
m+2\right)  \frac{P_{m+2}^{(n-m-2,0)}(\tau)}{P_{m}^{(n-m,0)}(\tau)},\label{26}%
\end{equation}
and%
\begin{equation}
\left \langle b^{\dagger2}b^{2}\right \rangle =\left(  n+1\right)  \left(
n+2\right)  \sinh^{4}\lambda \frac{P_{m}^{(n-m+2,0)}(\tau)}{P_{m}%
^{(n-m,0)}(\tau)},\label{27}%
\end{equation}

For the state $|\left \vert \lambda,m,n\right \rangle $, substituting
Eqs.(\ref{23}), (\ref{26})and (\ref{27}) into Eq.(\ref{25}), we can recast
$R_{ab}$\ to%

\begin{equation}
R_{ab}=\frac{\left(  m+1\right)  \left(  m+2\right)  P_{m+2}^{(n-m-2,0)}%
(\tau)+\left(  n+1\right)  \left(  n+2\right)  \sinh^{4}\lambda P_{m}%
^{(n-m+2,0)}(\tau)}{2\left(  m+1\right)  \left(  n+1\right)  \sinh^{2}\lambda
P_{m+1}^{(n-m,0)}(\tau)}-1. \label{28}%
\end{equation}
In particular, when $m=n=0$ (corresponding to two-mode squeezed vacuum state),
Eq.(\ref{28}) reduces to $R_{ab,m=n=0}=-\operatorname{sech}2\lambda<0,$ which
indicates that there always exist antibunching effect for two-mode squeezed
vacuum state. In addition, when $m=n,$ the TPSSV\textit{ }is always
antibunching. However, for any parameter values $m,n(m\neq n)$, the
case\textit{ }is not true. The $R_{ab}$ as a function of $\lambda$ and $m,n$
is plotted in Fig. 4. It is easy to see that, for a given $m$ the
TPSSV\textit{ }presents the antibunching effect when the squeezing parameter
$\lambda$ exceeds to a certain threshold value. For instance, when $m=0\ $and
$n=2$ then $R_{ab}=\frac{5-3\cosh2\lambda}{6\left(  1+2\cosh2\lambda \right)
}$csch$^{2}\lambda$ may be less than zero with $\lambda>0.549$ about.

\begin{figure}[ptb]
\label{Fig8.3}
\centering \includegraphics[width=15cm]{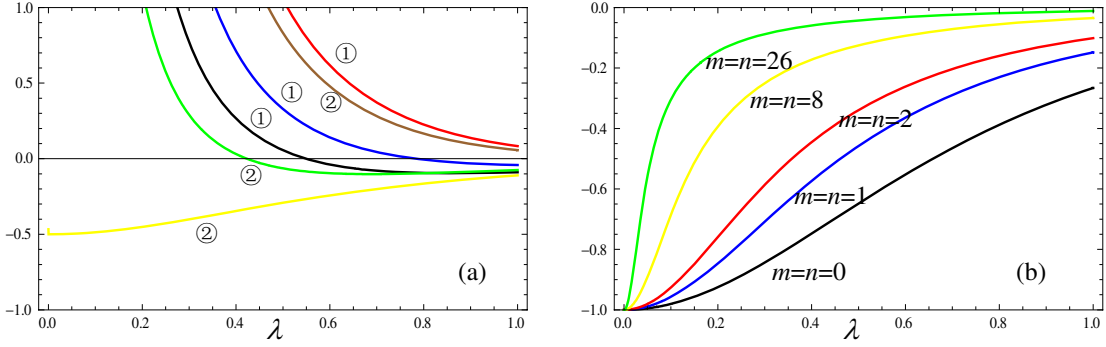}\caption{(Color online)
The $R_{ab}$ as a function of $\lambda$ and $m,n$. (1) and (2) in Fig.4 (a)
denote$m=0$ and $m=1$£¬respectively, and $n=2,3,12$ from down to up. }%
\end{figure}

\section{Wigner function of the TPSSV}

The Wigner function (WF)\cite{25,34,35} is a powerful tool to investigate the
nonclassicality of optical fields. Its partial negativity implies the highly
nonclassical properties of quantum states and is often used to describe the
decoherence of quantum states, e.g., the excited coherent state in both
photon-loss and thermal channels \cite{36,37}, the single-photon subtracted
squeezed vacuum (SPSSV) state in both amplitude decay and phase damping
channels \cite{2d}, and so on \cite{4,10,38,39,40}. In this section, we derive
the analytical expression of WF for the TPSSV. For this purpose, we first
recall that the Weyl ordered form of single-mode Wigner operator
\cite{41,42,43},%
\begin{equation}
\Delta_{1}\left(  \alpha \right)  =\frac{1}{2}%
\genfrac{}{}{0pt}{}{\colon}{\colon}%
\delta \left(  \alpha-a\right)  \delta \left(  \alpha^{\ast}-a^{\dag}\right)
\genfrac{}{}{0pt}{}{\colon}{\colon}%
,\label{29}%
\end{equation}
where $\alpha=\left(  q_{1}+ip_{1}\right)  /\sqrt{2}$ and the symbol $%
\genfrac{}{}{0pt}{}{\colon}{\colon}%
\genfrac{}{}{0pt}{}{\colon}{\colon}%
$ denotes Weyl ordering. The merit of Weyl ordering lies in the Weyl ordered
operators' invariance under similar transformations proved in Ref.\cite{41},
which means%
\begin{equation}
S%
\genfrac{}{}{0pt}{}{:}{:}%
\left(  \circ \circ \circ \right)
\genfrac{}{}{0pt}{}{:}{:}%
S^{-1}=%
\genfrac{}{}{0pt}{}{:}{:}%
S\left(  \circ \circ \circ \right)  S^{-1}%
\genfrac{}{}{0pt}{}{:}{:}%
,\label{30}%
\end{equation}
as if the \textquotedblleft fence" $%
\genfrac{}{}{0pt}{}{:}{:}%
\genfrac{}{}{0pt}{}{:}{:}%
$did not exist, so $S$ can pass through it.

Following this invariance and Eq.(\ref{3}) we have%
\begin{align*}
&  S_{2}^{\dag}\left(  \lambda \right)  \Delta_{1}\left(  \alpha \right)
\Delta_{2}\left(  \beta \right)  S_{2}\left(  \lambda \right)  \\
&  =\frac{1}{4}S_{2}^{\dag}\left(  \lambda \right)
\genfrac{}{}{0pt}{}{\colon}{\colon}%
\delta \left(  \alpha-a\right)  \delta \left(  \alpha^{\ast}-a^{\dag}\right)
\delta \left(  \beta-b\right)  \delta \left(  \beta^{\ast}-b^{\dag}\right)
\genfrac{}{}{0pt}{}{\colon}{\colon}%
S_{2}\left(  \lambda \right)  \\
&  =\frac{1}{4}%
\genfrac{}{}{0pt}{}{\colon}{\colon}%
\delta \left(  \alpha-a\cosh \lambda-b^{\dagger}\sinh \lambda \right)
\delta \left(  \alpha^{\ast}-a^{\dagger}\cosh \lambda-b\sinh \lambda \right)  \\
&  \delta \left(  \beta-b\cosh \lambda-a^{\dagger}\sinh \lambda \right)
\delta \left(  \beta^{\ast}-b^{\dagger}\cosh \lambda-a\sinh \lambda \right)
\genfrac{}{}{0pt}{}{\colon}{\colon}%
\\
&  =\frac{1}{4}%
\genfrac{}{}{0pt}{}{\colon}{\colon}%
\delta \left(  \bar{\alpha}-a\right)  \delta \left(  \bar{\alpha}^{\ast}%
-a^{\dag}\right)  \delta \left(  \bar{\beta}-b\right)  \delta \left(  \bar
{\beta}^{\ast}-b^{\dag}\right)
\genfrac{}{}{0pt}{}{\colon}{\colon}%
\\
&  =\Delta_{1}\left(  \bar{\alpha}\right)  \Delta_{2}\left(  \bar{\beta
}\right)  ,
\end{align*}
where $\bar{\alpha}=\alpha \cosh \lambda-\beta^{\ast}\sinh \lambda,$ $\bar{\beta
}=\beta \cosh \lambda-\alpha^{\ast}\sinh \lambda,$ and $\beta=\left(
q_{2}+ip_{2}\right)  /\sqrt{2}$. Thus employing the squeezed two-variable
Hermite-excited vacuum state of the TPSSV in Eq.(\ref{5}) and the coherent
state representation of single-mode Wigner operator \cite{44},%
\begin{equation}
\Delta_{1}(\alpha)=e^{2\left \vert \alpha \right \vert ^{2}}\int \frac{d^{2}z_{1}%
}{\pi^{2}}\left \vert z_{1}\right \rangle \left \langle -z_{1}\right \vert
e^{-2(z_{1}\alpha^{\ast}-\alpha z_{1}^{\ast})},\label{34}%
\end{equation}
where $\left \vert z_{1}\right \rangle =\exp \left(  z_{1}a^{\dag}-z_{1}^{\ast
}a\right)  \left \vert 0\right \rangle $ is Glauber coherent state \cite{30,31},
we finally can obtain the explicit expression of WF for the TPSSV (see
Appendix A),
\begin{align}
W(\alpha,\beta) &  =\frac{1}{\pi^{2}}\frac{\sinh^{n+m}2\lambda}{2^{n+m}%
N_{\lambda,m,n}}e^{-2\left \vert \bar{\alpha}\right \vert ^{2}-2\left \vert
\bar{\beta}\right \vert ^{2}}\sum_{l=0}^{m}\sum_{k=0}^{n}\nonumber \\
&  \times \frac{\left[  m!n!\right]  ^{2}\left(  -\tanh \lambda \right)  ^{l+k}%
}{l!k!\left[  \left(  m-l\right)  !\left(  n-k\right)  !\right]  ^{2}%
}\left \vert H_{m-l,n-k}\left(  B,A\right)  \right \vert ^{2},\label{35}%
\end{align}
where we have set $A=-2i\bar{\alpha}\sqrt{\tanh \lambda},B=-2i\bar{\beta}%
\sqrt{\tanh \lambda}.$ Obviously, the WF $W(\alpha,\beta)$ in Eq.(\ref{35}) is
a real function and is non-Gaussian in phase space due to the presence of
$H_{m-l,n-k}\left(  B,A\right)  $, as expected.

In particular, when $m=n=0,$ Eq.(\ref{35}) reduces to $W(\alpha,\beta
)=\frac{1}{\pi^{2}}e^{-2\left \vert \bar{\alpha}\right \vert ^{2}-2\left \vert
\bar{\beta}\right \vert ^{2}}=\frac{1}{\pi^{2}}e^{2\left(  \alpha^{\ast}%
\beta^{\ast}+\alpha \allowbreak \beta \right)  \sinh2\lambda-2\left(
\alpha \alpha^{\ast}+\beta \beta^{\ast}\allowbreak \right)  \cosh2\lambda}$
corresponding to the WF of two-mode squeezed vacuum state; while for the case
of $m=0$ and $n\neq0,$ noticing $H_{0,n}\left(  x,y\right)  =y^{n}$ and
$N_{\lambda,0,n}=n!\sinh^{2n}\lambda,$ Eq.(\ref{35}) becomes%
\begin{equation}
W(\alpha,\beta)=\frac{\left(  -1\right)  ^{n}}{\pi^{2}}e^{-2\left \vert
\bar{\alpha}\right \vert ^{2}-2\left \vert \bar{\beta}\right \vert ^{2}}%
L_{n}\left(  4\left \vert \bar{\alpha}\right \vert ^{2}\right)  ,\label{36}%
\end{equation}
where $L_{n}$ is $m$-order Laguerre polynomials and Eq.(\ref{36}) is just the
WF of the negative binomial state $S_{2}(\lambda)\left \vert n,0\right \rangle $
\cite{28}. In Figs. 5-7, the phase space Wigner distributions are depicted for
several different parameter values $m,n$, and $\lambda$. As an evidence of
nonclassicality of the state, squeezing in one of quadratures is clear in the
plots. In addition, there are some negative region of the WF in the phase
space which is another indicator of the nonclassicality of the state. For the
case of $m=0$ and $n=1,$ it is easily seen from (\ref{36}) that at the center
of the phase space ($\alpha=\beta=0$), the WF is always negative in phase
space. Fig.5 shows that the negative region becomes more and visible as the
increasement of photon number subtracted $m(=n)$, which may imply the
nonclassicality of the state can be enhanced due to the augment of
photon-subtraction number. For a given value $m$\ and several different values
$n$ ($\neq m$), the WF distributions are presented in Fig.7, from which it is
interseting to notice that there are around $\left \vert m-n\right \vert $ wave
valleys and $\left \vert m-n\right \vert +1$ wave peaks. \begin{figure}[ptb]
\label{Fig8.4}
\centering \includegraphics[width=12cm]{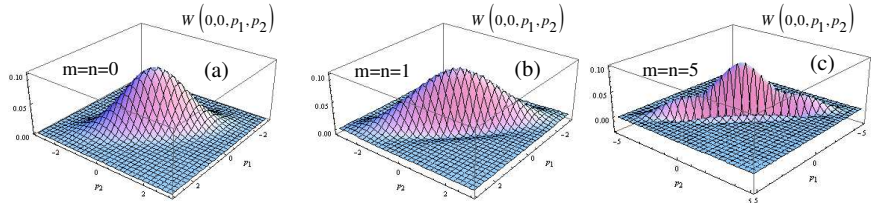}\caption{(Color online)
The Wigner function W($\alpha,\beta$) in phase space ($0,0,p_{1},p_{2}$) for
several different parameter values $m=n$ with $\lambda=0.5.$ (a) m=n=0; (b)
m=n=1 and (c) m=n=5.}%
\end{figure}\begin{figure}[ptb]
\label{Fig8.5}
\centering \includegraphics[width=12cm]{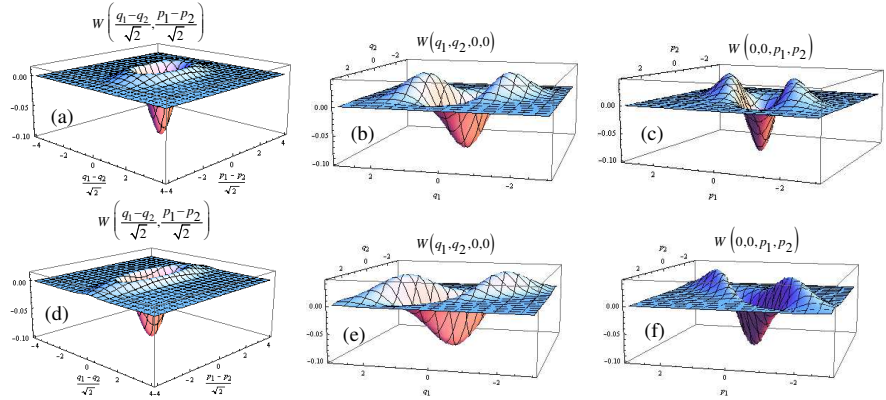}\caption{(Color online)
The Wigner function W($\alpha,\beta$) in three different phase spaces for
$m=0,n=1$ with $\lambda=0.3$ (first row) and $\lambda=0.5\ $(second low).}%
\end{figure}\begin{figure}[ptb]
\label{Fig8.6}
\centering \includegraphics[width=12cm]{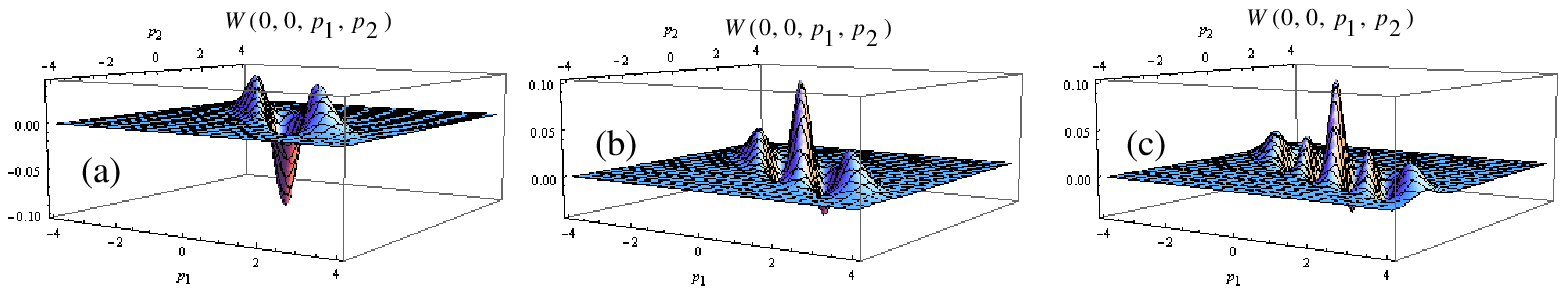}\caption{(Color online)
The Wigner function W($\alpha,\beta$) in phase space ($0,0,p_{1},p_{2}$) for
several parameter values $m,n$ with $\lambda=0.5.$ (a) m1,=n=2; (b) m=1,n=3
and (c) m=1,n=5.}%
\end{figure}

\section{Decoherence of TPSSV in thermal environments}

In this section, we next consider how this state evolves at the
presence of thermal environment.

\subsection{Model}

When the TPSSV evolves in the thermal channel, the evolution of the density
matrix can be described by the following master equation in the interaction
picture\cite{45}£º%
\begin{equation}
\frac{d}{dt}\rho \left(  t\right)  =\left(  L_{1}+L_{2}\right)  \rho \left(
t\right)  ,\label{37}%
\end{equation}
where%
\begin{align}
L_{i}\rho &  =\kappa \left(  \bar{n}+1\right)  \left(  2a_{i}\rho a_{i}^{\dag
}-a_{i}^{\dag}a_{i}\rho-\rho a_{i}^{\dag}a_{i}\right)  \nonumber \\
&  +\kappa \bar{n}\left(  2a_{i}^{\dag}\rho a_{i}-a_{i}a_{i}^{\dag}\rho-\rho
a_{i}a_{i}^{\dag}\right)  ,\text{ }\left(  a_{1}=a,a_{2}=b\right)  ,\label{38}%
\end{align}
and $\kappa$ represents the dissipative coefficient and $\bar{n}$
denotes the average thermal photon number of the environment. When
$\bar{n}=0,$ Eq.(\ref{37}) reduces to the master equation (ME)
describing the photon-loss channel. The two thermal modes are
assumed to have the same average energy and coupled with the channel
in the same strength and have the same average thermal photon number
$\bar{n}$. This assumption is reasonable as the two-mode of squeezed
state are in the same frequency and temperature of the environment
is normally the same \cite{46,47}. By introducing two entangled
state representations and using the technique of integration within
an ordered product (IWOP) of operators, we can obtain the infinite
operator-sum
expression of density matrix in Eq.(\ref{37}) (see Appendix B):%
\begin{equation}
\rho \left(  t\right)  =\sum_{i,j,r,s=0}^{\infty}M_{i,j,r,s}\rho_{0}%
M_{i,j,r,s}^{\dag},\label{39}%
\end{equation}
where $\rho_{0}$ denotes the density matrix at initial time, $M_{i,j,r,s}$ and
$M_{i,j,r,s}^{\dag}$ are Hermite conjugated operators (Kraus operator) with
each other,
\begin{equation}
M_{i,j,r,s}=\frac{1}{\bar{n}T+1}\sqrt{\frac{\left(  T_{1}\right)
^{r+s}\left(  T_{3}\right)  ^{i+j}}{r!s!i!j!}}a^{\dagger r}b^{\dagger
s}e^{\left(  a^{\dagger}a+b^{\dagger}b\right)  \ln T_{2}}a^{i}b^{j},\label{40}%
\end{equation}
and we have set $T=1-e^{-2\kappa t}$, as well as
\begin{equation}
T_{1}=\frac{\bar{n}T}{\bar{n}T+1},T_{2}=\frac{e^{-\kappa t}}{\bar{n}T+1}%
,T_{3}=\frac{\left(  \bar{n}+1\right)  T}{\bar{n}T+1}.\label{41}%
\end{equation}
It is not difficult to prove the $M_{i,j,r,s}$ obeys the
normalization condition
$\sum_{i,j,r,s=0}^{\infty}M_{i,j,r,s}^{\dag}M_{i,j,r,s}=1$ by using
the IWOP technique.

\subsection{Evolution of Wigner function}

By using the thermal field dynamics theory \cite{48,49} and thermal entangled
state representation, the time evolution of Wigner function at time $t$ to be
given by the convolution of the Wigner function at initial time and those of
two single-mode thermal state (see Appendix C), i.e.,%
\begin{equation}
W\left(  \alpha,\beta,t\right)  =\frac{4}{\left(  2\bar{n}+1\right)  ^{2}%
T^{2}}\int \frac{d^{2}\zeta d^{2}\eta}{\pi^{2}}W\left(  \zeta,\eta,0\right)
e^{-2\frac{\left \vert \alpha-\zeta e^{-\kappa t}\right \vert ^{2}+\left \vert
\beta-\eta e^{-\kappa t}\right \vert ^{2}}{\left(  2\bar{n}+1\right)  T}%
}.\label{42}%
\end{equation}
Eq.(\ref{42}) is just the evolution formula of Wigner function of two-mode
quantum state in thermal channel. Thus the WF at any time can be obtained by
performing the integration when the initial WF is known.

In a similar way to deriving Eq.(\ref{35}), substituting Eq.(\ref{35}) into
Eq.(\ref{42}) and using the generating function of two-variable Hermite
polynomials (A2), we finally obtain%

\begin{align}
W\left(  \alpha,\beta,t\right)   &  =\frac{N_{\lambda,m,n}^{-1}\left(
E\sinh2\lambda \right)  ^{m+n}}{\pi^{2}2^{n+m}\left(  2\bar{n}+1\right)
^{2}T^{2}D}e^{-\frac{\left \vert \alpha-\beta^{\ast}\right \vert ^{2}%
}{e^{-2\lambda-2\kappa t}+\left(  2\bar{n}+1\right)  T}-\frac{\left \vert
\alpha+\beta^{\ast}\right \vert ^{2}}{e^{2\lambda-2\kappa t}+\left(  2\bar
{n}+1\right)  T}}\nonumber \\
&  \times \sum_{l=0}^{n}\sum_{k=0}^{m}\frac{\left[  m!n!\right]  ^{2}\left(
-\frac{F}{E}\tanh \lambda \right)  ^{l+k}}{l!k!\left[  \left(  m-k\right)
!\left(  n-l\right)  !\right]  ^{2}}\left \vert H_{m-k,n-l}\left(  G/\sqrt
{E},K/\sqrt{E}\right)  \right \vert ^{2}, \label{43}%
\end{align}
where we have set%

\begin{align}
C  &  =\frac{e^{-2\kappa t}}{\left(  2\bar{n}+1\right)  T},\text{ }D=\left(
1+Ce^{-2\lambda}\right)  \left(  1+Ce^{2\lambda}\right)  ,\nonumber \\
E  &  =\allowbreak \frac{e^{4\kappa t}}{D}\left(  2\bar{n}T+1\right)  ^{2}%
C^{2},\text{ }F=\frac{C^{2}-1}{D},\nonumber \\
G  &  =\frac{Ce^{\kappa t}}{D}\left(  \bar{B}+B^{\ast}\allowbreak C\right)
,\text{ }\bar{B}=\allowbreak i2\sqrt{\tanh \lambda}\left(  \beta^{\ast}%
\cosh \lambda+\alpha \allowbreak \sinh \lambda \right)  ,\nonumber \\
K  &  =\frac{Ce^{\kappa t}}{D}\left(  \bar{A}+A^{\ast}C\right)  ,\text{ }%
\bar{A}=i2\sqrt{\tanh \lambda}\left(  \alpha^{\ast}\cosh \lambda+\beta
\sinh \lambda \right)  . \label{44}%
\end{align}
Eq.(\ref{43}) is just the analytical expression of WF for the TPSSV in thermal
channel. It is obvious that the WF loss its Gaussian property due to the
presence of two-variable Hermite polynomials.

In particular, at the initial time ($t=0$)$,$ noting $E\rightarrow1$, $\left(
2\bar{n}+1\right)  ^{2}T^{2}D\rightarrow1$, $\frac{F}{E}\rightarrow1$ and
$\frac{C^{2}}{D}\rightarrow1,$ $\frac{C}{D}\rightarrow0$ as well as
$K\rightarrow A^{\ast}$, $G\rightarrow B^{\ast}$, Eq.(\ref{43}) just dose
reduce to Eq.(\ref{35}), i.e., the WF of the TPSSV. On the other hand, when
$\kappa t\rightarrow \infty,$ noticing that $C\rightarrow0,D\rightarrow
1,E\rightarrow1,F\rightarrow-1,$ and $G/\sqrt{E}\rightarrow0,K/\sqrt
{E}\rightarrow0,$ as well as $H_{m,n}\left(  0,0\right)  =\left(  -1\right)
^{m}m!\delta_{m,n},$ as well as the definition of Jacobi polynomials in
Eq.(\ref{9}), then Eq.(\ref{43}) becomes%
\begin{equation}
W\left(  \alpha,\beta,\infty \right)  =\frac{1}{\pi^{2}\left(  2\bar
{n}+1\right)  ^{2}}e^{-\frac{2}{2\bar{n}+1}(\left \vert \alpha \right \vert
^{2}+\left \vert \beta \right \vert ^{2})},\label{45}%
\end{equation}
which is independent of photon-subtraction number $m$ and $n$ and
corresponds
to the product of two thermal states with mean thermal photon number $\bar{n}%
$. This implies that the two-mode system reduces to two-mode thermal state
after a long time interaction with the environment. Eq.(\ref{45}) denotes a
Gaussian distribution. Thus the thermal noise causes the absence of the
partial negative of the WF if the decay time $\kappa t$ exceeds a threshold
value. In addition, for the case of $m=n=0$, corresponding to the case of
two-mode squeezed vacuum, Eq.(\ref{43}) just becomes
\begin{equation}
W_{m=n=0}\left(  \alpha,\beta,t\right)  =\mathfrak{N}^{-1}e^{-\frac
{\mathfrak{E}}{\mathfrak{D}}\left(  \left \vert \alpha \right \vert
^{2}+\left \vert \beta \right \vert ^{2}\right)  +\frac{\mathfrak{F}%
}{\mathfrak{D}}\left(  \alpha \beta+\alpha^{\ast}\beta^{\ast}\right)
},\label{46}%
\end{equation}
where $\mathfrak{N}=\pi^{2}\left(  2\bar{n}+1\right)  ^{2}T^{2}D$ is the
normalization factor, $\mathfrak{D}=\left(  2\bar{n}+1\right)  ^{2}T^{2}D,$
$\mathfrak{E}=2\left(  2\bar{n}+1\right)  T+e^{-2\kappa t}\cosh2\lambda,$ and
$\mathfrak{F}=2e^{-2\kappa t}\sinh2\lambda$. Eq.(\ref{46}) is just the result
in Eq.(14) of Ref. \cite{47}.

In Fig.8, the WFs of the TPSSV for ($m=0,n=1$) are depicted in phase space
with $\lambda=0.3$ and $\bar{n}=1$ for several different $\kappa t.$ It is
easy to see that the negative region of WF gradually disappears as the time
$\kappa t$ increases. Actually, from Eq.(\ref{44}) one can see that $D>0$ and
$E>0$, so when $F<0$ leading to the following condition:%
\begin{equation}
\kappa t>\kappa t_{c}\equiv \frac{1}{2}\ln \frac{2\bar{n}+2}{2\bar{n}+1},
\label{47}%
\end{equation}
we know that the WF of TPSSV has no chance to be negative in the whole phase
space when $\kappa t\ $exceeds a threshold value $\kappa t_{c}$. Here we
should point out that the effective threshold value of the decay time
corresponding to the transition of the WF from partial negative to fully
positive definite is dependent of $m\ $and $n.$ When $\kappa t=\kappa t_{c},$
it then follows from Eq.(\ref{43}) that%
\begin{align}
W\left(  \alpha,\beta,t_{c}\right)   &  =\frac{\tanh^{m+n}\lambda
\operatorname{sech}^{2}\lambda}{4\pi^{2}N_{m,n,\lambda}e^{-4\kappa t_{c}}%
}e^{-e^{2\kappa t_{c}}\left[  \left \vert \alpha \right \vert ^{2}+\left \vert
\beta \right \vert ^{2}-\left(  \alpha^{\ast}\beta^{\ast}+\alpha \beta \right)
\tanh \lambda \right]  }\nonumber \\
&  \times \left \vert H_{m,n}\left(  i\sqrt{\tanh \lambda}\beta^{\ast}e^{\kappa
t_{c}},\allowbreak i\sqrt{\tanh \lambda}\alpha^{\ast}e^{\kappa t_{c}}\right)
\right \vert ^{2}, \label{48}%
\end{align}
which is an Hermite-Gaussian function and positive definite, as expected.

In Figs. 9 and 10, we have presented the time-evolution of WF in
phase space for different $\bar{n}$ and $\lambda,$ respectively. One
can see clearly that the partial negativity of WF decreases
gradually as $\bar{n}$ (or $\lambda$) increases for a given time.
This case is true for a given $\bar{n}$ (or $\kappa t$) as the
increasement of $\kappa t$ (or $\bar{n}$). The squeezing effect in
one of quadratures is shown in Fig.10. In principle, by using the
explicit expression of WF in Eq.(\ref{43}), we can draw its
distributions in phase space. For the case of $m=0,n=2$, there are
two negative regions of WF, which is different from the case of
$m=0,n=1$ (see Fig.11). The absolute value of the negative minimum
of the WF decreases as $\kappa t$ increases, which leads to the full
absence of partial negative region.\begin{figure}[ptb]
\label{Fig8.8} \centering
\includegraphics[width=12cm]{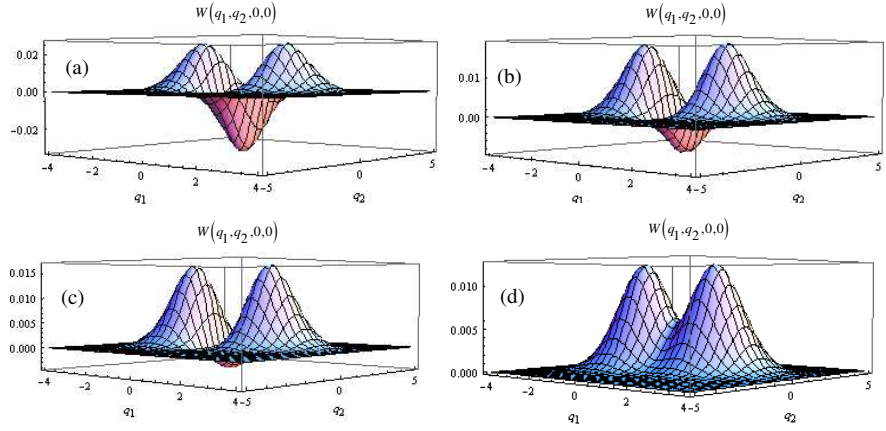}\caption{(Color online)
The time evolution of WF $\left(  m=0,n=1\right)  $ at $\left(  q_{1}%
,q_{2},0,0\right)  $\ phase space for $\bar{n}=1,\lambda=0.3.$(a) $\kappa
t=0.05,$(b) $\kappa t=0.1,$(c) $\kappa t=0.12,$(d) $\kappa t=0.2.$}%
\end{figure}\begin{figure}[ptb]
\label{Fig8.9}
\centering \includegraphics[width=12cm]{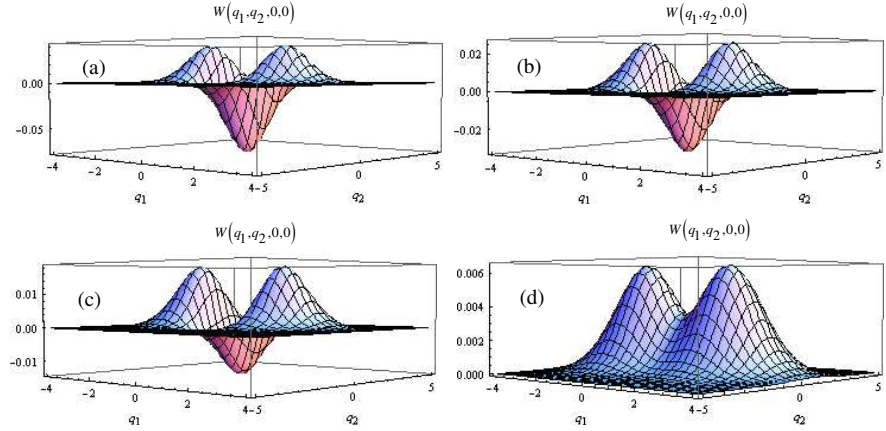}\caption{(Color online)
The time evolution of WF $\left(  m=0,n=1\right)  $ in $\left(  q_{1}%
,q_{2},0,0\right)  $\ phase space for $\lambda=0.3\ $and $\kappa t=0.05\ $with
(a) $\bar{n}=0,$(b) $\bar{n}=1,$(c) $\bar{n}=2,$(d) $\bar{n}=7.$}%
\end{figure}\begin{figure}[ptb]
\label{Fig8.10}
\centering \includegraphics[width=12cm]{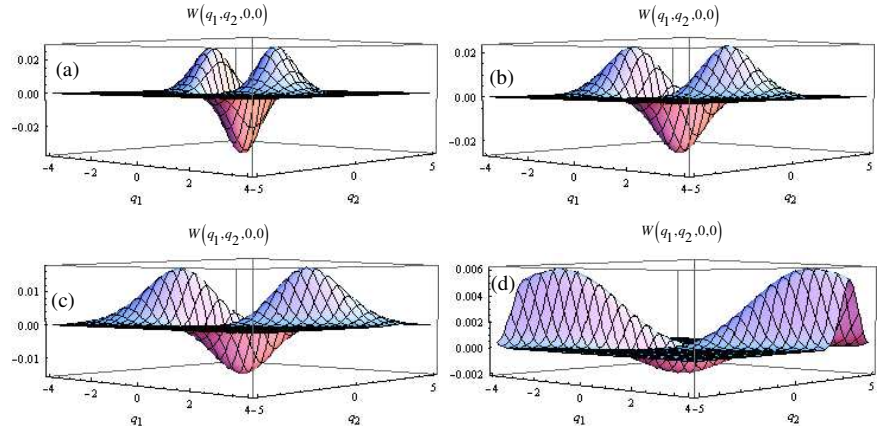}\caption{(Color online)
The time evolution of WF $\left(  m=0,n=1\right)  $ in $\left(  q_{1}%
,q_{2},0,0\right)  $\ phase space for $\bar{n}=1,$and $\kappa t=0.05\ $with
(a) $\lambda=0.03,$(b) $\lambda=0.5,$(c) $\lambda=0.8,$(d) $\lambda=1.2.$}%
\end{figure}\begin{figure}[ptb]
\label{Fig8.11}
\centering \includegraphics[width=12cm]{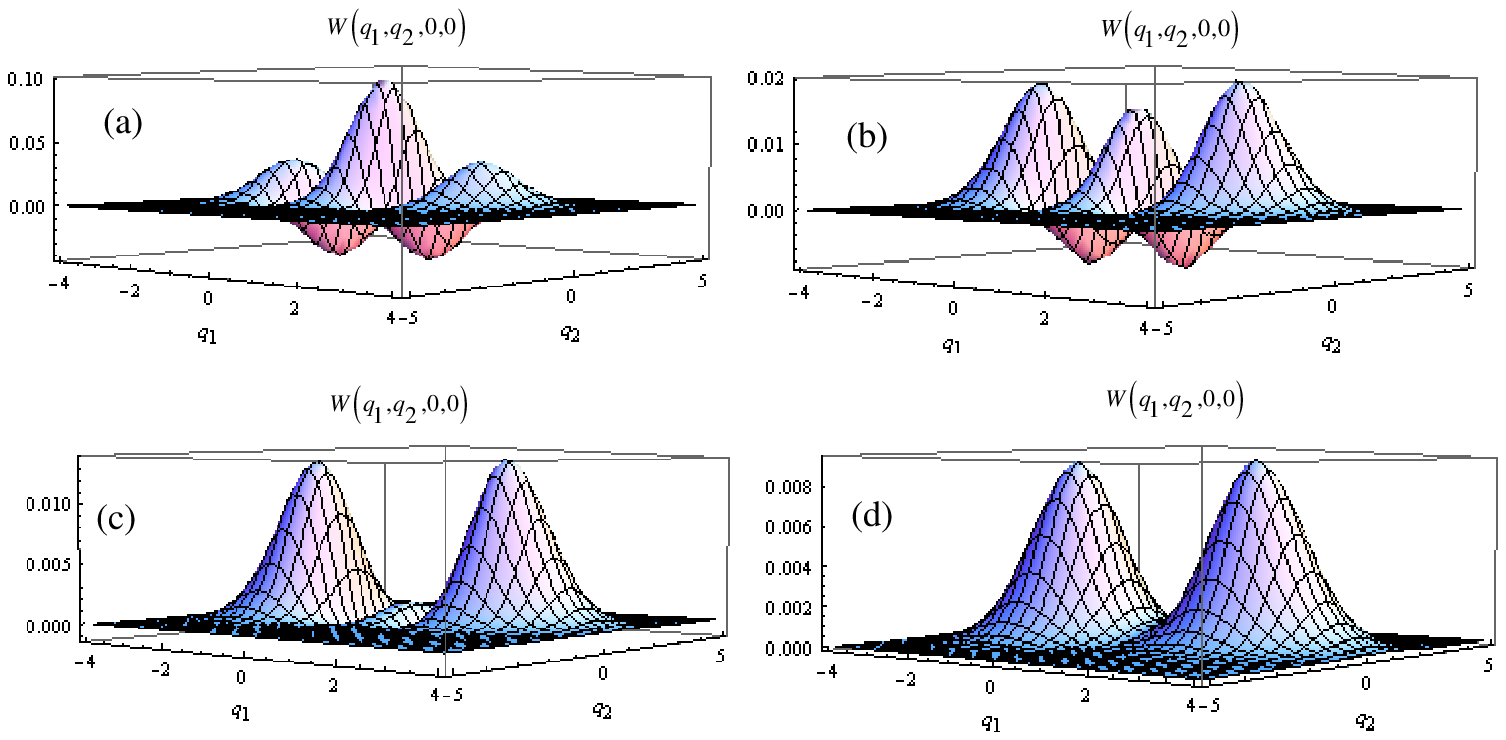}\caption{(Color online)
The time evolution of WF for $m=0,n=2$ in $\left(  q_{1},q_{2},0,0\right)
$\ phase space with (a) $\kappa t=0,$(b) $\kappa t=0.05,$(c) $\kappa
t=0.1,$(d) $\kappa t=0.2.$}%
\end{figure}

\section{Conclusions}

In summary, we have investigated the statistical properties of
two-mode photon-subtracted squeezed vacuum state (TPSSV) and its
decoherence in thermal channelwith average thermal photon number
$\bar{n}$\ and dissipative coefficient $\kappa$. For arbitrary
number TPSSV, we have for the first time calculated the
normalization factor, which turns out to be a Jacobi polynomial of
the squeezing parameter $\lambda$, a remarkable result. We also show
that the TPSSV can be treated as a squeezed two-variable Hermite
polynomial excitation vacuum. Based on Jacobi polynomials' behavior
the statistical properties of the field, such as photon number
distribution, squeezing properties, cross-correlation function and
antibunching, are also derived analytically. Especially, the
nonclassicality of TPSSV is discussed in terms of the negativity of
WF after deriving the explicit expression of WF. Then the
decoherence of TPSSV in thermal channel is also demonstrated
according to the compact expression for the WF. The threshold value
of the decay time corresponding to the transition of the WF from
partial negative to completely positive is presented. It is found
that the WF has no chance to present negative value for all
parameters $\lambda$ and any photon-subtraction number ($m,n$) if
$\kappa t>\frac{1}{2}\ln \frac{2\bar{n}+2}{2\bar{n}+1}\ $for TPSSV.
The technique of integration within an ordered product of operators
brings convenience in our derivation.

\textbf{Acknowledgments }Work supported by the the National Natural Science
Foundation of China under Grant Nos.10775097 and 10874174.

\textbf{Appendix A: Deriviation of Wigner function Eq.(\ref{35}) of TPSSV}

The definite of the WF of two-mode quantum state $\left \vert \Psi \right \rangle
$ is given by $W(\alpha,\beta)=\left \langle \Psi \right \vert \Delta_{1}\left(
\alpha \right)  \Delta_{2}\left(  \beta \right)  \left \vert \Psi \right \rangle $,
thus by uisng Eqs.(\ref{5}), and (\ref{34}) the WF of TPSSV can be calculated
as%
\begin{align}
W(\alpha,\beta)  &  =\left \langle \lambda,m,n\right \vert |\Delta_{1}\left(
\alpha \right)  \Delta_{2}\left(  \beta \right)  |\left \vert \lambda
,m,n\right \rangle \nonumber \\
&  =\frac{\sinh^{n+m}2\lambda}{2^{n+m}N_{\lambda,m,n}}\left \langle
00\right \vert H_{m,n}\left(  -i\sqrt{\tanh \lambda}b,-i\sqrt{\tanh \lambda
}a\right)  \Delta_{1}\left(  \bar{\alpha}\right) \nonumber \\
&  \otimes \Delta_{2}\left(  \bar{\beta}\right)  H_{m,n}\left(  i\sqrt
{\tanh \lambda}b^{\dagger},i\sqrt{\tanh \lambda}a^{\dagger}\right)  \left \vert
00\right \rangle \nonumber \\
&  =\frac{\sinh^{n+m}2\lambda}{2^{n+m}N_{\lambda,m,n}}e^{2\left \vert
\bar{\alpha}\right \vert ^{2}+2\left \vert \bar{\beta}\right \vert ^{2}}\int
\frac{d^{2}z_{1}d^{2}z_{2}}{\pi^{4}}e^{-\left \vert z_{1}\right \vert
^{2}-\left \vert z_{2}\right \vert ^{2}-2(z_{1}\bar{\alpha}^{\ast}-\bar{\alpha
}z_{1}^{\ast})-2(z_{2}\bar{\beta}^{\ast}-\bar{\beta}z_{2}^{\ast})}\nonumber \\
&  \times H_{m,n}\left(  -i\sqrt{\tanh \lambda}z_{2},-i\sqrt{\tanh \lambda}%
z_{1}\right)  H_{m,n}\left(  -i\sqrt{\tanh \lambda}z_{2}^{\ast},-i\sqrt
{\tanh \lambda}z_{1}^{\ast}\right)  . \tag{A1}%
\end{align}
Further noticing the generating function of two variables Hermitian
polynomials,
\begin{equation}
H_{m,n}\left(  \epsilon,\varepsilon \right)  =\frac{\partial^{m+n}}{\partial
t^{m}\partial t^{\prime n}}\left.  \exp \left[  -tt^{\prime}+\epsilon
t+\varepsilon t^{\prime}\right]  \right \vert _{t=t^{\prime}=0}, \tag{A2}%
\end{equation}
Eq.(A1) can be further rewritten as%
\begin{align}
W(\alpha,\beta)  &  =\frac{\sinh^{n+m}2\lambda}{2^{n+m}N_{\lambda,m,n}%
}e^{2\left \vert \bar{\alpha}\right \vert ^{2}+2\left \vert \bar{\beta
}\right \vert ^{2}}\frac{\partial^{m+n}}{\partial t^{m}\partial \tau^{n}}%
\frac{\partial^{m+n}}{\partial t^{\prime m}\partial \tau^{\prime n}}%
e^{-t\tau-t^{\prime}\tau^{\prime}}\nonumber \\
&  \times \int \frac{d^{2}z_{1}}{\pi^{2}}\left.  e^{-\left \vert z_{1}\right \vert
^{2}+\left(  -2\bar{\alpha}^{\ast}-i\sqrt{\tanh \lambda}\tau \right)
z_{1}+\left(  2\bar{\alpha}-i\sqrt{\tanh \lambda}\tau^{\prime}\right)
z_{1}^{\ast}}\right \vert _{t=\tau=0}\nonumber \\
&  \times \int \frac{d^{2}z_{2}}{\pi^{2}}\left.  e^{-\left \vert z_{2}\right \vert
^{2}+\left(  -2\bar{\beta}^{\ast}-i\sqrt{\tanh \lambda}t\right)  z_{2}+\left(
2\bar{\beta}-i\sqrt{\tanh \lambda}t^{\prime}\right)  z_{2}^{\ast}}\right \vert
_{t^{\prime}=\tau^{\prime}=0}\nonumber \\
&  =\frac{\sinh^{n+m}2\lambda}{2^{n+m}N_{\lambda,m,n}}e^{-2\left \vert
\bar{\alpha}\right \vert ^{2}-2\left \vert \bar{\beta}\right \vert ^{2}}%
\frac{\partial^{m+n}}{\partial t^{m}\partial \tau^{n}}\frac{\partial^{m+n}%
}{\partial t^{\prime m}\partial \tau^{\prime n}}\nonumber \\
&  \times \left.  e^{-t\tau-t^{\prime}\tau^{\prime}+A^{\ast}\tau \prime+B^{\ast
}t^{\prime}\allowbreak+A\tau+Bt-\left(  tt^{\prime}+\tau \tau^{\prime}\right)
\tanh \lambda}\right \vert _{t=\tau=t^{\prime}=\tau^{\prime}=0}, \tag{A3}%
\end{align}
where we have set
\begin{equation}
B=-2i\bar{\beta}\sqrt{\tanh \lambda},A=-2i\bar{\alpha}\sqrt{\tanh \lambda},
\tag{A4}%
\end{equation}
and have used the following integration formula
\begin{equation}
\int \frac{d^{2}z}{\pi}e^{\zeta \left \vert z\right \vert ^{2}+\xi z+\eta z^{\ast
}}=-\frac{1}{\zeta}e^{-\frac{\xi \eta}{\zeta}},\text{Re}\left(  \zeta \right)
<0. \tag{A5}%
\end{equation}

Expanding the exponential term $\exp \left[  -\left(  tt^{\prime}+\tau
\tau^{\prime}\right)  \tanh \lambda \right]  ,$ and using Eq.(A2), we have%
\begin{align}
W(\alpha,\beta) &  =\frac{\sinh^{n+m}2\lambda}{2^{n+m}N_{\lambda,m,n}%
}e^{-2\left \vert \bar{\alpha}\right \vert ^{2}-2\left \vert \bar{\beta
}\right \vert ^{2}}\sum_{l=0}^{\infty}\sum_{k=0}^{\infty}\frac{\left(
-\tanh \lambda \right)  ^{l+k}}{l!k!}\nonumber \\
&  \allowbreak \times \frac{\partial^{l+k}}{\partial B^{l}\partial A^{k}}%
\frac{\partial^{l+k}}{\partial B^{\ast l}\partial A^{\ast k}}\frac
{\partial^{2m}}{\partial t^{m}\partial \tau^{n}}\nonumber \\
&  \times \left.  \frac{\partial^{2n}}{\partial t^{\prime m}\partial
\tau^{\prime n}}e^{-t\tau+A\tau+Bt-t^{\prime}\tau^{\prime}+A^{\ast}\tau
\prime+B^{\ast}t^{\prime}\allowbreak}\right \vert _{t=\tau=t^{\prime}%
=\tau^{\prime}=0}\nonumber \\
&  =\frac{\sinh^{n+m}2\lambda}{2^{n+m}N_{\lambda,m,n}}e^{-2\left \vert
\bar{\alpha}\right \vert ^{2}-2\left \vert \bar{\beta}\right \vert ^{2}}%
\sum_{l=0}^{\infty}\sum_{k=0}^{\infty}\frac{\left(  -\tanh \lambda \right)
^{l+k}}{l!k!}\nonumber \\
&  \times \frac{\partial^{l+k}}{\partial B^{l}\partial A^{k}}\frac
{\partial^{l+k}}{\partial B^{\ast l}\partial A^{\ast k}}H_{m,n}\left(
B,A\right)  H_{m,n}\left(  B^{\ast},A^{\ast}\right)  .\tag{A6}%
\end{align}
Noticing the well-known differential relations of $H_{m,n}\left(
\epsilon,\varepsilon \right)  ,$
\begin{equation}
\frac{\partial^{l+k}}{\partial \epsilon^{l}\partial \varepsilon^{k}}%
H_{m,n}\left(  \epsilon,\varepsilon \right)  =\frac{m!n!H_{m-l,n-k}\left(
\epsilon,\varepsilon \right)  }{\left(  m-l\right)  !\left(  n-k\right)
!},\tag{A7}%
\end{equation}
we can further recast Eq.(A6) to Eq.(\ref{35}).

\textbf{Appendix B: Derivation of solution of Eq.(\ref{37})}

To solve the ME in Eq.(\ref{37}), we first introduce two entangled state
representations \cite{49a}:
\begin{align}
\left \vert \eta_{a}\right \rangle  &  =\exp \left[  -\frac{1}{2}|\eta_{a}%
|^{2}+\eta_{a}a^{\dagger}-\eta_{a}^{\ast}\tilde{a}^{\dagger}+a^{\dagger}%
\tilde{a}^{\dagger}\right]  \left \vert 0\tilde{0}\right \rangle ,\tag{B1}\\
\left \vert \eta_{b}\right \rangle  &  =\exp \left[  -\frac{1}{2}|\eta_{b}%
|^{2}+\eta_{b}b^{\dagger}-\eta_{b}^{\ast}\tilde{b}^{\dagger}+b^{\dagger}%
\tilde{b}^{\dagger}\right]  \left \vert 0\tilde{0}\right \rangle ,\tag{B2}%
\end{align}
which satisfy the following eigenvector equations, for instance,%
\begin{equation}%
\begin{array}
[c]{c}%
(a-\tilde{a}^{\dagger})\left \vert \eta_{a}\right \rangle =\eta_{a}\left \vert
\eta_{a}\right \rangle ,\;(a^{\dagger}-\tilde{a})\left \vert \eta_{a}%
\right \rangle =\eta_{a}^{\ast}\left \vert \eta_{a}\right \rangle ,\\
\left \langle \eta_{a}\right \vert (a^{\dagger}-\tilde{a})=\eta_{a}^{\ast
}\left \langle \eta_{a}\right \vert ,\  \left \langle \eta_{a}\right \vert
(a-\tilde{a}^{\dagger})=\eta_{a}\left \langle \eta_{a}\right \vert .
\end{array}
\tag{B3}%
\end{equation}
which imply operators $(a-\tilde{a}^{\dagger})$ and $(a^{\dagger}-\tilde{a})$
can be replaced by number $\eta_{a}$ and$\  \eta_{a}^{\ast},$ $\left[
(a-\tilde{a}^{\dagger}),(a^{\dagger}-\tilde{a})\right]  =0.$ Operating
two-side of Eq.(\ref{37}) on the vector $\left \vert I_{a},I_{b}\right \rangle
\equiv \left \vert \eta_{a}=0\right \rangle \otimes \left \vert \eta_{b}%
=0\right \rangle $, (denote $\left \vert \rho \left(  t\right)  \right \rangle
\equiv \rho \left(  t\right)  \left \vert I_{a},I_{b}\right \rangle ),$ and
noticing the corresponding relation:%
\begin{equation}%
\begin{array}
[c]{c}%
a\left \vert I_{a},I_{b}\right \rangle =\tilde{a}^{\dagger}\left \vert
I_{a},I_{b}\right \rangle ,\text{ }a^{\dagger}\left \vert I_{a},I_{b}%
\right \rangle =\tilde{a}\left \vert I_{a},I_{b}\right \rangle ,\\
b\left \vert I_{a},I_{b}\right \rangle =\tilde{b}^{\dagger}\left \vert
I_{a},I_{b}\right \rangle ,\text{ }b^{\dagger}\left \vert I_{a},I_{b}%
\right \rangle =\tilde{b}\left \vert I_{a},I_{b}\right \rangle ,
\end{array}
\tag{B4}%
\end{equation}
we can put Eq.(\ref{37}) into the following form:%
\begin{align}
\frac{d}{dt}\left \vert \rho \left(  t\right)  \right \rangle  &  =\left[
\kappa \left(  \bar{n}+1\right)  \left(  2a\tilde{a}-a^{\dag}a-\tilde
{a}^{\dagger}\tilde{a}\right)  +\kappa \bar{n}\left(  2a^{\dag}\tilde
{a}^{\dagger}-aa^{\dag}-\tilde{a}\tilde{a}^{\dag}\right)  \right.  \nonumber \\
&  \left.  +\kappa \left(  \bar{n}+1\right)  \left(  2b\tilde{b}-b^{\dag
}b-\tilde{b}^{\dag}\tilde{b}\right)  +\kappa \bar{n}\left(  2b^{\dag}\tilde
{b}^{\dag}-bb^{\dag}-\tilde{b}\tilde{b}^{\dag}\right)  \right]  \left \vert
\rho \left(  t\right)  \right \rangle .\tag{B5}%
\end{align}
It's formal solution is given by
\begin{align}
\left \vert \rho \left(  t\right)  \right \rangle  &  =\exp \left[  \kappa
t\left(  \bar{n}+1\right)  \left(  2a\tilde{a}-a^{\dag}a-\tilde{a}^{\dagger
}\tilde{a}\right)  +\kappa t\bar{n}\left(  2a^{\dag}\tilde{a}^{\dagger
}-aa^{\dag}-\tilde{a}\tilde{a}^{\dag}\right)  \right.  \nonumber \\
&  \left.  +\kappa t\left(  \bar{n}+1\right)  \left(  2b\tilde{b}-b^{\dag
}b-\tilde{b}^{\dag}\tilde{b}\right)  +\kappa t\bar{n}\left(  2b^{\dag}%
\tilde{b}^{\dag}-bb^{\dag}-\tilde{b}\tilde{b}^{\dag}\right)  \right]
\left \vert \rho_{0}\right \rangle ,\tag{B6}%
\end{align}
where $\left \vert \rho_{0}\right \rangle \equiv \rho_{0}\left \vert I_{a}%
,I_{b}\right \rangle $. In order to solve Eq.(B6), noticing that, for example,
\begin{equation}
2a\tilde{a}-a^{\dag}a-\tilde{a}^{\dagger}\tilde{a}=-\left(  a^{\dag}-\tilde
{a}\right)  \left(  a-\tilde{a}^{\dag}\right)  +\tilde{a}a-\tilde{a}^{\dagger
}a^{\dag},\tag{B7}%
\end{equation}
we have%
\begin{align}
\left \vert \rho \left(  t\right)  \right \rangle  &  =\exp \left[  \left(
a\tilde{a}-\tilde{a}^{\dagger}a^{\dagger}+1\right)  \kappa t\right]
\nonumber \\
&  \times \exp \left[  \frac{2\bar{n}+1}{2}\left(  1-e^{2\kappa t}\right)
\left(  a^{\dagger}-\tilde{a}\right)  \left(  a-\tilde{a}^{\dagger}\right)
\right]  \nonumber \\
&  \times \exp \left[  \left(  b\tilde{b}-\tilde{b}^{\dagger}b^{\dagger
}+1\right)  \kappa t\right]  \nonumber \\
&  \times \exp \left[  \frac{2\bar{n}+1}{2}\left(  1-e^{2\kappa t}\right)
\left(  b^{\dagger}-\tilde{b}\right)  \left(  b-\tilde{b}^{\dagger}\right)
\right]  \left \vert \rho_{0}\right \rangle ,\tag{B8}%
\end{align}
where we have used the identity operator, $\exp[\lambda(A+\sigma
B)]=e^{\lambda A}\exp[\sigma B(1-e^{-\lambda \tau})/\tau]$ valid for
$[A,B]=\tau B.$

Thus the element of $\rho \left(  t\right)  $ between $\left \langle \eta
_{a},\eta_{b}\right \vert $ and $\left \vert I_{a},I_{b}\right \rangle $ is%
\begin{equation}
\left \langle \eta_{a},\eta_{b}\right \vert \left.  \rho \left(  t\right)
\right \rangle =\exp \left[  -\frac{2\bar{n}+1}{2}T|\left(  \eta_{a}|^{2}%
+|\eta_{b}|^{2}\right)  \right]  \left \langle \eta_{a}e^{-\kappa t},\eta
_{b}e^{-\kappa t}\right \vert \left.  \rho_{0}\right \rangle ,\tag{B9}%
\end{equation}
from which one can see clearly the attenuation due to the presence of environment.

Further, using the completeness relation of $\left \vert \eta_{a},\eta
_{b}\right \rangle $, $\int \frac{d^{2}\eta_{a}d^{2}\eta_{b}}{\pi^{2}}\left \vert
\eta_{a},\eta_{b}\right \rangle \left \langle \eta_{a},\eta_{b}\right \vert =1$
and the IWOP technique \cite{50,51}, we see%
\begin{align}
\left \vert \rho \left(  t\right)  \right \rangle  &  =\int \frac{d^{2}\eta
_{a}d^{2}\eta_{b}}{\pi^{2}}\left \vert \eta_{a},\eta_{b}\right \rangle
\left \langle \eta_{a},\eta_{b}\right \vert \left.  \rho \left(  t\right)
\right \rangle \nonumber \\
&  =\frac{1}{\left(  \bar{n}T+1\right)  ^{2}}\exp \left[  T_{1}\left(
a^{\dagger}\tilde{a}^{\dagger}+b^{\dagger}\tilde{b}^{\dagger}\right)  \right]
\nonumber \\
&  \times \exp \left[  \left(  a^{\dagger}a+b^{\dagger}b+\tilde{a}^{\dagger
}\tilde{a}+\tilde{b}^{\dagger}\tilde{b}\right)  \ln T_{2}\right]  \nonumber \\
&  \times \exp \left[  T_{3}\left(  a\tilde{a}+b\tilde{b}\right)  \right]
\rho_{0}\left \vert I_{a},I_{b}\right \rangle ,\tag{B10}%
\end{align}
where $T_{1},T_{2}$ and $T_{3}$ are defined in Eq.(\ref{41}). Noticing
Eq.(B4), we can reform Eq.(B10) as $\rho \left(  t\right)  =\sum_{i,j,r,s=0}%
^{\infty}M_{i,j,r,s}\rho_{0}M_{i,j,r,s}^{\dag},$where $M_{i,j,r,s}$ and
$M_{i,j,r,s}^{\dag}$ are defined in Eq.(\ref{40}).

\textbf{Appendix C: Deriviation of Eq.(\ref{42}) by using thermo field
dynamics and entangled state representation}

In this appendix, we shall derive the evolution formula of WF, i.e.,
the relation between the any time WF and the initial time WF.
According to the definition of WF of density operator $\rho$:
$W\left(  \alpha \right) =\mathtt{Tr}\left[  \Delta \left(  \alpha
\right)  \rho \right]  $, where $\Delta \left(  \alpha \right)  $ is
the single-mode Wigner operator, $\Delta \left(  \alpha \right)
=\frac{1}{\pi}D\left(  2\alpha \right)  \left( -1\right)
^{a^{\dag}a}$. By using $\left \langle \tilde{n}\right.  \left \vert
\tilde{m}\right \rangle =\delta_{m,n}$ we can reform $W\left(
\alpha \right)  $
as \cite{52}%
\begin{equation}
W\left(  \alpha \right)  =\sum_{m,n}^{\infty}\left \langle n,\tilde
{n}\right \vert \Delta \left(  \alpha \right)  \rho \left \vert m,\tilde
{m}\right \rangle =\frac{1}{\pi}\left \langle \xi_{=2\alpha}\right \vert \left.
\rho \right \rangle ,\tag{C1}%
\end{equation}
where $\left \langle \xi \right \vert $ is the conjugate state of $\left \langle
\eta \right \vert $, whose overlap is $\left \langle \eta \right \vert \left.
\xi \right \rangle =\frac{1}{2}\exp \left[  \frac{1}{2}\left(  \xi \eta^{\ast}%
-\xi^{\ast}\eta \right)  \right]  ,$a Fourier transformation kernel. In a
similar way, thus for two-mode quantum system, the WF is given by%
\begin{equation}
W\left(  \alpha,\beta \right)  =\mathtt{Tr}\left[  \Delta_{a}\left(
\alpha \right)  \Delta_{b}\left(  \beta \right)  \rho \right]  =\frac{1}{\pi^{2}%
}\left \langle \xi_{a=2\alpha},\xi_{b=2\beta}\right \vert \left.  \rho
\right \rangle .\tag{C2}%
\end{equation}
Employing the above overlap relation, Eq.(C2) can be recast into the following
form:%
\begin{equation}
W\left(  \alpha,\beta,t\right)  =\int \frac{d^{2}\eta_{a}d^{2}\eta_{b}}%
{4\pi^{4}}e^{\alpha^{\ast}\eta_{a}-\alpha \eta_{a}^{\ast}+\beta^{\ast}\eta
_{b}-\beta \eta_{b}^{\ast}}\left \langle \eta_{a},\eta_{b}\right \vert \left.
\rho \left(  t\right)  \right \rangle .\tag{C3}%
\end{equation}
Then substituting Eq.(B9) into Eq.(C3) and using the completeness of
$\left \langle \xi \right \vert $, $\int \frac{d^{2}\xi}{\pi}\left
\vert \xi \right \rangle \left \langle \xi \right \vert =1,$ we have
\begin{align}
W\left(  \alpha,\beta,t\right)   &  =\int \frac{d^{2}\eta_{a}d^{2}\eta_{b}%
}{4\pi^{4}}e^{-\frac{2\bar{n}+1}{2}T|\left(  \eta_{a}|^{2}+|\eta_{b}%
|^{2}\right)  }\nonumber \\
&  \times e^{\alpha^{\ast}\eta_{a}-\alpha \eta_{a}^{\ast}+\beta^{\ast}\eta
_{b}-\beta \eta_{b}^{\ast}}\left \langle \eta_{a}e^{-\kappa t},\eta
_{b}e^{-\kappa t}\right \vert \left.  \rho_{0}\right \rangle \nonumber \\
&  =\int \frac{d^{2}\xi_{a}d^{2}\xi_{b}}{\pi^{2}}W\left(  \zeta,\eta,0\right)
\int \frac{d^{2}\eta_{a}d^{2}\eta_{b}}{4\pi^{2}}e^{-\frac{2\bar{n}+1}%
{2}T|\left(  \eta_{a}|^{2}+|\eta_{b}|^{2}\right)  }\nonumber \\
&  \times e^{\alpha^{\ast}\eta_{a}-\alpha \eta_{a}^{\ast}+\beta^{\ast}\eta
_{b}-\beta \eta_{b}^{\ast}}\left \langle \eta_{a}e^{-\kappa t},\eta
_{b}e^{-\kappa t}\right \vert \left.  \xi_{a=2\zeta},\xi_{b=2\eta}\right \rangle
.\tag{C4}%
\end{align}
Performing the integration in Eq.(C4) over $d^{2}\eta_{a}d^{2}\eta_{b}$ then
we can obtain Eq.(\ref{42}).

Making variables replacement, $\frac{\alpha-\zeta e^{-\kappa t}}{\sqrt{T}%
}\rightarrow \zeta,$ $\frac{\beta-\eta e^{-\kappa t}}{\sqrt{T}}\rightarrow
\eta,$ Eq.(\ref{42}) can be reformed as
\begin{align}
W\left(  \alpha,\beta,t\right)   &  =4e^{4\kappa t}\int d^{2}\zeta d^{2}\eta
W_{a}^{th}\left(  \zeta \right)  W_{b}^{th}\left(  \eta \right)  \nonumber \\
&  \times W\left \{  e^{\kappa t}\left(  \alpha-\sqrt{T}\zeta \right)
,e^{\kappa t}\left(  \beta-\sqrt{T}\eta \right)  ,0\right \}  ,\tag{C5}%
\end{align}
where $W^{th}\left(  \zeta \right)  $ is the Wigner function of
thermal state with average thermal photon number $\bar{n}$:
$W^{th}\left(  \zeta \right) =\frac{1}{\pi \left(  2\bar{n}+1\right)
}e^{-\frac{2\left \vert \zeta \right \vert ^{2}}{2\bar{n}+1}}.$
Eq.(C5) is\ another expression of the evolution of WF and is
actually agreement with that in Refs.\cite{46,47}.

\end{document}